\documentclass[english]{revtex4}
\usepackage[T1]{fontenc}
\usepackage[latin9]{inputenc}
\usepackage{graphicx}

\providecommand{\tabularnewline}{\\}

\usepackage{babel}
\ifx\ave\undefined

\global\long\global\long\def\ave#1{\left\langle #1 \right\rangle }

\fi

\begin{document}

\title{On the energy dependence of $K/\pi$ fluctuations in relativistic
heavy ion collisions}

\author{Volker Koch}

\email{vkoch@lbl.gov}

\affiliation{Nuclear Science Division\\
 Lawrence Berkeley National Laboratory\\
 Berkeley, CA, 94720, USA}

\author{Tim Schuster}

\email{Tim.Schuster@cern.ch}

\affiliation{Department of Physics\\
 University of Frankfurt\\
 Frankfurt, Germany}

\begin{abstract}
In this note we will discuss the energy dependence of particle ratio
fluctuations in heavy ion collisions. We study how the inherent multiplicity
dependence of ratio fluctuations is reflected in the excitation function
of the dynamical fluctuations. Specifically, we will show that the
observed excitation function of dynamical $K/\pi$ fluctuations is
consistent with the expected dependence on the number of accepted
pions and kaons in both the STAR and NA49 experiments. 
\end{abstract}
\maketitle

\section{introduction}

Over the last several years event-by-event fluctuations of many observables
have been studied in relativistic heavy ion collisions~\cite{Afanasev:2000fu,Alt:2004ir,Alt:2008ca,Grebieszkow:2007xz,Rybczynski:2008cv,Abelev:2008jg,Abelev:2009if,Adamova:2003pz,Adams:2003st,Adams:2003uw,Adcox:2002mm,Adcox:2002pa,Adler:2003xq}.
The measurement of these fluctuations may reveal separate event classes,
fluctuations associated with phase-transitions etc.\ (for a recent review
see~\cite{Koch:2008ia}). The first measurement of event-by-event
hadron ratio fluctuations has been carried out by the NA49 collaboration, which
analyzed the fluctuations of the kaon to pion ($K/\pi)$ ratio in
lead-lead collisions at a center of mass energy of $\sqrt{s}=17.3\:{\rm GeV}$~\cite{Afanasev:2000fu}.
This measurement found surprisingly little fluctuations, of the order
of a few percent, which could be explained by a combination of Bose-Einstein
correlations together with resonance decays~\cite{Jeon:1999gr,Stephanov:1999zu}.

Subsequently, the NA49 collaboration has measured event-by-event ratio fluctuations
for several values of the center of mass energy, ranging from $\sqrt{s}=6.3\,{\rm GeV}$
to $\sqrt{s}=17.3\,{\rm GeV}$~\cite{Alt:2008ca}. In addition, at
the Relativistic Heavy Ion Collider (RHIC) the STAR, PHENIX, and PHOBOS
collaborations have measured fluctuations up to center of mass energies
of $\sqrt{s}=200\,{\rm GeV}$ so that excitation functions of many
fluctuation observables are now available over a wide range of energies.
While most excitation functions show only little energy dependence,
that for the kaon to pion ratio exhibits a steep increase towards
the lower energies. So far this increase could not be reproduced in
either thermal model calculations~\cite{Torrieri:2007vv} nor with
the microscopic transport model UrQMD~\cite{Alt:2008ca} and thus
has sparked quite some interest and speculations concerning the QCD
phase transition~\cite{Koch:2005pk}. Another calculation using the HSD
event-generator \cite{Gorenstein:2008et} can describe the general increase of
the fluctuations towards lower energies but fails to reproduce the very steep
rise exhibited in the NA49 data.

In this note, we will point out that the observable generally used
in the discussion of dynamical fluctuations,\begin{equation}
\sigma_{{\rm dynamical}}^{2}=\sigma_{{\rm data}}^{2}-\sigma_{{\rm mixed\ events}}^{2}\label{eq:sig_dynamic}\end{equation}
 has an inherent dependence on the multiplicity of particles that
are used in the experimental analysis - these are the particles located
within the phase space domain covered by experimental acceptance with
particle identification capability. This has some nontrivial consequences
for excitation functions extracted with fixed target experiments,
such as NA49, where the acceptance changes considerably with beam
energy. Specifically, in case of $K/\pi$ - ratio fluctuations, the
dynamical fluctuations depend on the inverse of the \emph{accepted}
number of pions and kaons, and since their number decreases with beam
energy this may lead to non-negligible corrections, as we shall show.
Indeed in~\cite{Abelev:2009if} a significant centrality dependence
of these ratio fluctuations was found at top RHIC energies, consistent
with an inverse multiplicity scaling. In addition for a fixed target experiment
the actual acceptance changes with beam energy. Since it is the multiplicity
of the \emph{accepted} particles which determines the fluctuations,
this needs to be taken into account if one studies an excitation function.

It is the purpose of this note to discuss the multiplicity dependence
of particle ratio fluctuations and in particular those of the kaon-to-pion
ratio. After a brief review of the underlying formalism governing
fluctuations of (particle) ratios, we will discuss several ways to
remove the inherent multiplicity dependence in the definition of the
dynamical fluctuations, $\sigma_{{\rm dynamical}}^{2}$, Eq.~\ref{eq:sig_dynamic}.
We then apply them to the fluctuations of the kaon-to-pion ratio and
discuss the energy dependence of this observable in this context.

\section{Fluctuations of Particle Ratios\label{sec:Fluctuations-of-Particle}}

Following~\cite{Jeon:1999gr,Koch:2008ia} the fluctuations of a particle
ratio $A/B$ are given to leading order by\begin{equation}
\sigma^{2}=\frac{\ave{\left(\delta\frac{A}{B}\right)^{2}}}{\ave{\frac{A}{B}}^{2}
}=\left(\frac{\ave{\delta A^{2}}}{\ave A^{2}}+\frac{\ave{\delta B^{2}}}{\ave
B^{2}}-2\frac{\ave{\delta A\,\delta B}}{\ave A\ave B}\right)
+\mathcal{O}(\delta^4)
\label{eq:variance}
\end{equation}
 with the definition\begin{eqnarray*}
\delta A & = & A-\ave A\\
\delta B & = & B-\ave B\end{eqnarray*}
 Since\begin{eqnarray*}
\ave{\delta A^{2}} & = & \ave{A^{2}}-\ave A^{2}\\
\ave{\delta B^{2}} & = & \ave{B^{2}}-\ave B^{2}\\
\ave{\delta A\,\delta B} & = & \ave{A\, B}-\ave A\ave B\end{eqnarray*}
 Eq.~\ref{eq:variance} can also be written as\[
\sigma^{2}=\left(\frac{\ave{A^{2}}}{\ave A^{2}}+\frac{\ave{B^{2}}}{\ave B^{2}}-2\frac{\ave{A\, B}}{\ave A\ave B}\right)\]
 In absence of any correlation, $\ave{\delta A^{2}}=\ave A$ and $\ave{\delta B^{2}}=\ave B$,
and the scaled variance (Eq.~\ref{eq:variance}) reduces to\begin{equation}
\sigma_{{\rm uncorrelated}}^{2}=\left(\frac{1}{\ave A}+\frac{1}{\ave B}\right).\label{eq:variance_uncorr}\end{equation}
 The difference between the scaled variance $\sigma^{2}$ and the uncorrelated
scaled variance $\sigma_{{\rm uncorrelated}}^{2}$ is commonly referred to
as the dynamical fluctuations, $\sigma_{{\rm dynamical}}^{2}$,\begin{eqnarray}
\sigma_{{\rm dynamical}}^{2} & = & \sigma^{2}-\sigma_{{\rm uncorrelated}}^{2}\nonumber \\
 & = & \left(\frac{\ave{\delta A^{2}}-\ave A}{\ave A^{2}}+\frac{\ave{\delta B^{2}}-\ave B}{\ave B^{2}}-2\frac{\ave{\delta A\,\delta B}}{\ave A\ave B}\right)\nonumber \\
 & = & \left(\frac{\ave{A\left(A-1\right)}}{\ave A^{2}}+\frac{\ave{B\left(B-1\right)}}{\ave B^{2}}-2\frac{\ave{A\, B}}{\ave A\ave B}\right)\nonumber \\
 & = & \nu_{{\rm dynamical}}\label{eq:sig_dyn_explicit}\end{eqnarray}
 where $\nu_{{\rm dynamical}}$ is the variable usually used by the
STAR collaboration. Obviously in the absence of any correlations,
$\sigma_{{\rm {dynamical}}}=0$, by construction. Introducing the
scaled correlations\begin{equation}
C_{AB}\equiv\frac{\ave{\delta A\delta B}-\delta_{AB}\ave A}{\sqrt{\ave A\ave B}}\label{eq:scaled_corr}\end{equation}
 the dynamical fluctuations, $\sigma_{{\rm dynamical}}^{2}$,
can be written as\begin{equation}
\sigma_{{\rm dynamical}}^{2}=\left(\frac{1}{\ave A}C_{AA}+\frac{1}{\ave B}C_{BB}-\frac{2}{\sqrt{\ave A\ave B}}C_{AB}\right)\label{eq:sig_dyn_scal}\end{equation}
 We note that the scaled correlations, $C_{AB}$, typically do not
or only weakly depend on the multiplicity as we will show explicitly
in the context of a resonance gas. Consequently, Eq.~\ref{eq:sig_dyn_scal}
already suggests, that a simple scaling of $\sigma_{{\rm dynamical}}^{2}$
with number of charged particles may not be sufficient. Depending
on which of the scaled correlations dominates, $\sigma_{{\rm dynamical}}^{2}$
may either scale with $1/\ave A$, $1/\ave B$ or with $1/\sqrt{\ave A\ave B}$
or some combination of those. Furthermore, if the particle abundances
differ considerably, say $\frac{\ave A}{\ave B}\ll1$, as it is the
case for the kaon to pion ratio, the dynamical fluctuations will be
dominated by the least abundant particle, even if the scaled correlations
are of the same magnitude. This follows directly from Eq.~\ref{eq:sig_dyn_scal}.
In this case a scaling with $1/\ave A$ should work best.

Finally, quantum statistics gives rise to additional correlations
\cite{Mrowczynski:1998vt,Baym:1999up,Jeon:1999gr} \begin{equation}
\ave{\delta A^{2}}\simeq\ave A\left(1\pm\frac{\ave{n_{A}^{2}}}{\ave{n_{A}}}\right)\label{eq:bose_einstein}\end{equation}
 with\[
\ave{n_{A}^{2}}=\int\frac{{\rm d}^{3}p}{\left(2\pi\right)^{3}}\left[
n_{A}(p)\right]^2 \]
 where $\left(+\right)$ stands for Bosons and $\left(-\right)$ for
Fermions and $n_{A}(p)\,{\rm d}p$ is the number of particles of type
$A$ in momentum bin $\left(p,p+{\rm d}p\right)$. The correction
term due to quantum statistics is typically of the order of a $5-10\%$
for systems of consideration, resulting in ${\cal O}\left(1-2\%\right)$
effects for the dynamical fluctuations~\cite{Mrowczynski:1998vt,Jeon:1999gr}.
After these general remarks about fluctuations of particle ratios
and their scaling with multiplicity let us turn to the specific case
of kaon-to-pion ratio fluctuations.

\section{Multiplicity Scaling of $K/\pi$ fluctuations}

Let us now turn to the specific case of $K/\pi$ fluctuations. In
this case the scaled variance, Eq.~\ref{eq:variance}, is given by
\begin{equation}
\sigma_{K/\pi}^{2}=\frac{\ave{\left(\delta\frac{K}{\pi}\right)^{2}}}{\ave{\frac{
K}{\pi}}^{2}}=\left(\frac{\ave{(\delta K)^{2}}}{\ave
K^{2}}+\frac{\ave{(\delta\pi)^{2}}}{\ave{\pi}^{2}}-2\frac{\ave{\delta
K\,\delta\pi}}{\ave K\ave{\pi}}\right)\label{eq:variance_kp}
\end{equation}
 and the dynamical fluctuations, Eq.~\ref{eq:sig_dyn_scal}, are 
\[
\sigma_{{\rm dynamical}}^{2}=\left(\frac{1}{\ave K}C_{KK}+\frac{1}{\ave{\pi}}C_{\pi\pi}-\frac{2}{\sqrt{\ave K\ave{\pi}}}C_{K\pi}\right)\]
 Since $K=K_{+}+K_{-}$ and $\pi=\pi_{+}+\pi_{-}$

\begin{eqnarray}
\ave K & = & \ave{K_{+}+K_{-}}\nonumber \\
\ave{\pi} & = & \ave{\pi_{+}+\pi_{-}}\\
\ave{(\delta K)^{2}} & = & \ave{(\delta K_{+})^{2}}+\ave{(\delta
K_{-})^{2}}+2\ave{\delta K_{+}\delta K_{-}}\\
\ave{(\delta\pi)^{2}} & = &
\ave{(\delta\pi_{+})^{2}}+\ave{(\delta\pi_{-})^{2}}+2\ave{\delta\pi_{+}
\delta\pi_ { - } }\nonumber \\
\ave{\delta K\,\delta\pi} & = & \ave{\delta K_{+}\delta\pi_{+}}+\ave{\delta
K_{-}\delta\pi_{+}}+\ave{\delta K_{+}\delta\pi_{-}}+\ave{\delta
K_{-}\delta\pi_{-}}\label{eq:numerators}\end{eqnarray}
 so that the scaled correlations, Eq.~\ref{eq:scaled_corr}, are given
by
\begin{eqnarray}
C_{KK} & = & \frac{\ave{(\delta K_{+})^{2}}+\ave{(\delta
K_{-})^{2}}+2\ave{\delta K_{+}\delta
K_{-}}-\ave{K_{+}+K_{-}}}{\ave{K_{+}+K_{-}}}\nonumber \\
C_{\pi\pi} & = &
\frac{\ave{(\delta\pi_{+})^{2}}+\ave{(\delta\pi_{-})^{2}}+2\ave{\delta\pi_{+}
\delta\pi_{-}}-\ave{\pi_{+}+\pi_{-}}}{\ave{\pi_{+}+\pi_{-}}}\nonumber \\
C_{K\pi} & = & \frac{\ave{\delta K_{+}\delta\pi_{+}}+\ave{\delta
K_{-}\delta\pi_{+}}+\ave{\delta K_{+}\delta\pi_{-}}+\ave{\delta
K_{-}\delta\pi_{-}}}{\sqrt{\ave{K_{+}+K_{-}}\ave{\pi_{+}+\pi_{-}}}}
\label{eq:scal_corr_kp}
\end{eqnarray}
 We note, that the ``diagonal'' scaled correlations, $C_{KK}$
and $C_{\pi\pi}$, also contain cross-correlations between the positively
and negatively charged kaons and pions respectively. Therefore, correlations
introduced by resonances such as the $\phi$-meson and the $\rho_{0}$-meson
will \emph{enhance} the ``diagonal'' scaled correlations and thus the
dynamical fluctuations. Resonances decaying into a kaon and a pion,
such as the $K_{0}^{*}$-meson will contribute to the off-diagonal
scaled correlation, $C_{K\pi}$ and will reduce the dynamical fluctuations.
This will be different if charge specific ratios such as $\frac{\delta K_{+}}{\delta\pi_{-}}$
are considered. In this case only resonances which decay in either
two $K_{+}$-mesons or two $\pi_{-}$-mesons will contribute to the
diagonal terms, while the $K_{0}^{*}$-mesons do contribute to the
off-diagonal scaled correlation. Consequently, the fluctuations of
the charge specific ratio is not necessarily the same as that of the
ratio of sums of negative and positive kaons/pions. This is also seen
in the data by the STAR collaboration~\cite{Abelev:2009if}.

It may be instructive to discuss the above equations in the context
of a simple model system which contains $n_{\pi^{\pm}}$ charged pions
and $n_{K^{\pm}}$ charged kaons as well as $n_{\rho_{0}}$ \emph{neutral}
rho-mesons, $n_{\omega}$ omega-mesons, $n_{\phi}$ phi-mesons, and
$n_{K_{0}^{*}}$ neutral ${\rm K}_{0}^{*}$ and $n_{\bar{K}_{0}^{*}}$
$\bar{K}_{0}^{*}$-mesons. Via their decay channels these resonances
will give rise to the various correlation terms in Eq.~\ref{eq:scal_corr_kp}.
In reality, of course, there are many other resonances contributing
to $\sigma$ and to the scaled correlations. The ones chosen here
are the lightest and most abundant ones which lead to correlations
and thus should provide a rough idea on how the different terms contribute~\footnote{Here
we ignore the charged $K^{*}$-mesons. While they contribute
via the kaon and pion multiplicities, they do not introduce any correlations,
as their decay channels involve always a neutral kaon or pion. %
}. A detailed investigation would involve a full study in the hadron
resonance gas model, which is not the purpose of this paper. In addition
to correlations due to resonances there are effects due to quantum
statistics~\cite{Mrowczynski:1998vt,reif_book}, as already discussed.
Here, for simplicity we work with classical (Boltzmann) statistics.
Noting that the branching ratio $BR\left(\phi\rightarrow K_{+}+K_{-}\right)\simeq\frac{1}{2}$
and $BR\left(K_{0}^{*}\rightarrow K_{+}+\pi_{-}\right)=BR\left(\bar{K}_{0}^{*}\rightarrow K_{-}+\pi_{+}\right)\simeq\frac{2}{3}$
the average particle numbers in our simple model are given by\begin{eqnarray*}
\ave K & = &
\ave{K_{+}+K_{-}}=n_{K_{+}}+n_{K_{-}}+n_{\phi}+\frac{2}{3}\left(n_{K_{0}^{*}}+n_
{ \bar{K}_{0}^{*}}\right)\\
\ave{\pi} & = &
\ave{\pi_{+}+\pi_{-}}=n_{\pi_{+}}+n_{\pi_{-}}+2n_{\rho_{0}}+2n_{\omega}+\frac{2}
{ 3}\left(n_{K_{0}^{*}}+n_{\bar{K}_{0}^{*}}\right)\end{eqnarray*}
 and we obtain the following expression~\cite{Jeon:1999gr} for the
(co)-variances in Eq.~\ref{eq:numerators} 
\begin{eqnarray*}
\ave{(\delta K_{+})^{2}} & = &
n_{K_{+}}+\frac{1}{2} n_{\phi}+\frac{2}{3}n_{K_{0}^{*}}\\
\ave{(\delta K_{-})^{2}} & = &
n_{K_{-}}+ \frac{1}{2} n_{\phi} +\frac{2}{3}n_{\bar{K}_{0}^{*}}\\
\ave{(\delta\pi_{+})^{2}} & = &
n_{\pi_{+}}+n_{\rho_{0}}+n_{\omega}+\frac{2}{3}n_{\bar{K}_{0}^{*}}\\
\ave{(\delta\pi_{-})^{2}} & = &
n_{\pi_{-}}+n_{\rho_{0}}+n_{\omega}+\frac{2}{3}n_{K_{0}^{*}}\\
\ave{\delta K_{+}\delta K_{-}} & = & \frac{1}{2}n_{\phi}\\
\ave{\delta\pi_{+}\delta\pi_{-}} & = & n_{\rho_{0}} + n_{\omega}\\
\ave{\delta K_{+}\delta\pi_{-}} & = & \frac{2}{3}n_{K_{0}^{*}}\\
\ave{\delta K_{-}\delta\pi_{+}} & = & \frac{2}{3}n_{\bar{K}_{0}^{*}}
\end{eqnarray*}
 leading to (see Eq.~\ref{eq:numerators})
\begin{eqnarray*}
\ave{(\delta K)^{2}} & = & \ave{\delta K_{+}^{2}+\delta K_{-}^{2}+2\left(\delta
K_{+}\delta K_{-}\right)}=\ave K+n_{\phi}\\
\ave{(\delta\pi)^{2}} & = &
\ave{\delta\pi_{+}^{2}+\delta\pi_{-}^{2}+2\left(\delta\pi_{+}\delta\pi_{-}
\right)}=\ave{\pi}+2\left(n_{\rho_{0}}+n_{\omega}\right)\\
\ave{\delta K\,\delta\pi} & = & \frac{2}{3}\left(n_{K_{0}^{*}}+n_{\bar{K}_{0}^{*}}\right)\end{eqnarray*}
 The corresponding scaled correlations, Eq.~\ref{eq:scaled_corr},
are\begin{eqnarray}
C_{KK} & = & \frac{n_{\phi}}{\ave K}\nonumber \\
C_{\pi\pi} & = & \frac{2\left(n_{\rho_{0}}+n_{\omega}\right)}{\ave{\pi}}\nonumber \\
C_{K\pi} & = & \frac{2}{3}\frac{n_{K_{0}^{*}}+n_{\bar{K}_{0}^{*}}}{\sqrt{\ave K\ave{\pi}}}\label{eq:scaled_corr_model}\end{eqnarray}
 We find, that the scaled correlations in our simple model indeed
depend only weakly (if at all) on the multiplicity, since both the
number of resonances and the number of kaons and pions are expected
to scale roughly with the multiplicity or volume of the system. In
a thermal system at fixed temperature they would be constant. Furthermore,
as already discussed the correlations introduced via the resonances
affect all three scaled correlations. While the $K_{0}^{*}$-mesons
control the off-diagonal correlation term, $\ave{\delta K\,\delta\pi}$,
both the $\phi$-meson and the $\rho_{0}$ and $\omega$ contribute
to the diagonal parts. The former reduce the dynamical fluctuations
while the latter increase them. Putting everything together, the scaled
variance,
Eq.~\ref{eq:variance}, is given by\[
\sigma_{K/\pi}^{2}=\left(\frac{\ave K+2n_{\phi}}{\ave K^{2}}+\frac{\ave{\pi}+2\left(n_{\rho_{0}}+n_{\omega}\right)}{\ave{\pi}^{2}}-2\,\frac{\frac{2}{3}\left(n_{K_{0}^{*}}+n_{\bar{K}_{0}^{*}}\right)}{\ave K\ave{\pi}}\right)\]
 leading to \[
\sigma_{{\rm dynamical}}^{2}=\left(\frac{n_{\phi}}{\ave K^{2}}+\frac{2\left(n_{\rho_{0}}+n_{\omega}\right)}{\ave{\pi}^{2}}-2\,\frac{\frac{2}{3}\left(n_{K_{0}^{*}}+n_{\bar{K}_{0}^{*}}\right)}{\ave K\ave{\pi}}\right).\]
 or in terms of the scaled correlations\begin{equation}
\sigma_{{\rm dynamical}}^{2}=\left(\frac{1}{\ave K}C_{KK}+\frac{1}{\ave{\pi}}C_{\pi\pi}-\frac{2}{\sqrt{\ave K\ave{\pi}}}\, C_{K\pi}\right)\label{eq:sig_dyn_scaled_corr}\end{equation}
 Evaluating scaled correlations for our simple model, Eq.~\ref{eq:scaled_corr_model},
for a temperature of $T=170\,{\rm MeV}$ and vanishing chemical potential,
we get \begin{eqnarray*}
C_{KK} & = & 0.1\\
C_{\pi\pi} & = & 0.36\\
C_{K\pi} & = & 0.13\end{eqnarray*}
 Obviously, in our simple model all the scaled correlations are of
the same order of magnitude. In addition, adding more resonances will
likely reduce $C_{\pi\pi}$ as there are many resonances decaying
into only one pion, which add to the denominator, $\ave{\pi}$, but
not the numerator of $C_{\pi\pi}$. While it would be worthwhile to
study these scaled correlation coefficients in a full hadron gas model,
here instead we want to concentrate on simple phenomenological scaling
rules, with a special emphasis on the effect of a varying acceptance.

\section{\textbf{Phenomenological Scaling}}

In this section we want to discuss several ways to scale out the multiplicity
dependence of the dynamical fluctuations. Of course, if all the relevant
scaled correlations are known and if they, as we argued, depend only
weakly on the multiplicity and beam energy, the appropriate scaling
is simply given by Eq.~\ref{eq:sig_dyn_scaled_corr}. This is equivalent
to having full understanding of all the sources for the fluctuations
in which case this discussion is mute. In general, however, we do
not have a full understanding of all the sources contributing to the
fluctuations. In this case the multiplicity/energy dependence may
provide additional information such as signals for a possible phase
transition etc. In order to extract this information, one needs to
understand {}``trivial'' dependencies on the multiplicity, as exhibited
e.g.\ in Eq.~\ref{eq:sig_dyn_scaled_corr}. In the following we will
discuss several {}``trivial'' scaling prescriptions which we believe
should be applied before conclusions about new phenomena can be drawn.
We will focus on the rather interesting energy dependence of the $K/\pi$-fluctuations,
referring to~\cite{Abelev:2009if} for a discussion on the centrality
dependence. In addition to applying appropriate scaling prescriptions
to the dynamical fluctuations, it is essential to realize that the
multiplicities which control the fluctuations, such as $\ave K$ and
$\ave{\pi}$ are that of the \emph{identified} particles and \emph{not}
the extrapolated total multiplicities or mid-rapidity multiplicities.
This is especially important when studying the energy dependence of
fixed target data, as the acceptance varies with beam energy. We will
discuss this point in detail below after we have introduced the various
scaling prescriptions.

\subsection{Multiplicity Scaling Prescriptions}

\begin{figure}[ht]
\includegraphics[width=0.7\textwidth]{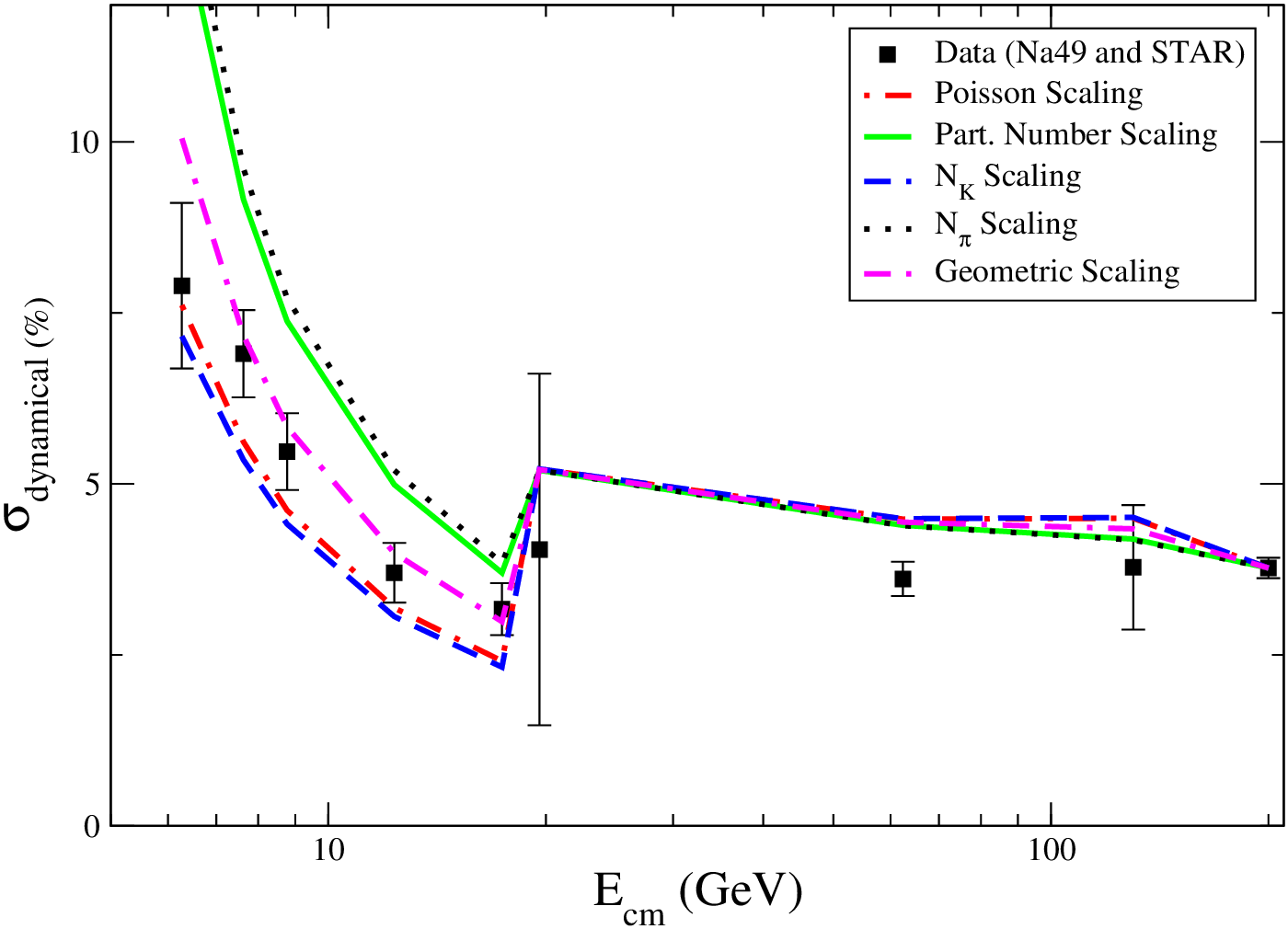}
\caption{Different scaling scenarios,
Eqs.~\ref{eq:poisson_scaling}-\ref{eq:geom_scaling}
based on the $\sqrt{s}=200\,{\rm GeV}$ data from STAR}

\label{fig:star_scale} 
\end{figure}

\begin{figure}[ht]
 \includegraphics[width=0.7\textwidth]{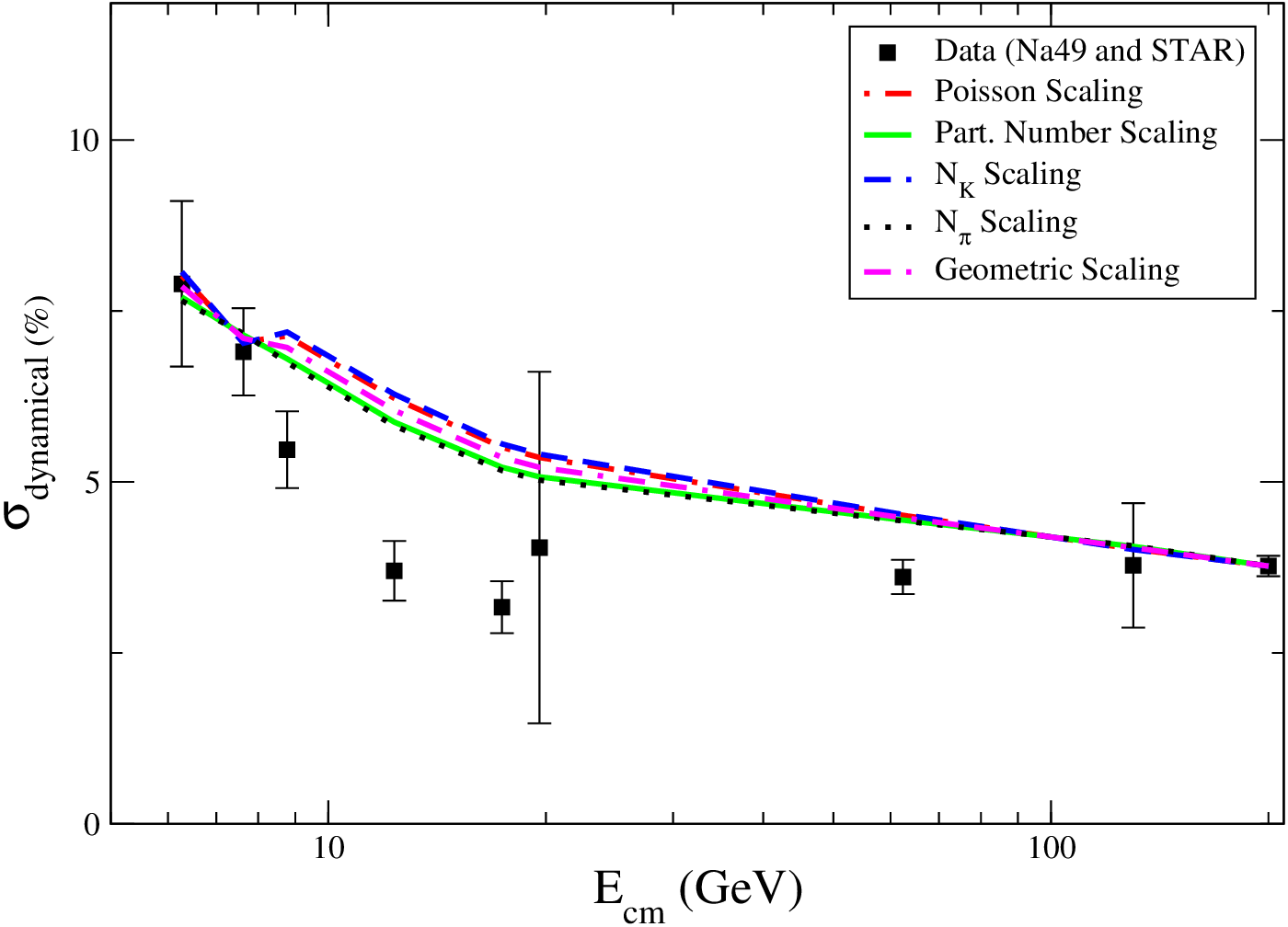}

\caption{Scaling with $\frac{{\rm d}N}{{\rm d}y}(y=0)$ instead of the actual number of
identified particles contributing to $\sigma_{{\rm dynamical}}$.
Otherwise same scaling formulae are used as in Fig.~\ref{fig:star_scale}. }

\label{fig:star_scale_dndy} 
\end{figure}

As suggested in~\cite{Jeon:1999gr}, one way to avoid any scaling
of the dynamical fluctuations with the number of accepted particles
would be to take the ratio of the measured scaled variance $\sigma^{2}$
over that of mixed events $\sigma_{{\rm mixed}}^{2}$, \[
f\equiv\frac{\sigma^{2}}{\sigma_{{\rm mixed}}^{2}},\]
instead of the difference as it is done in the definition of $\sigma_{{\rm dynamical}}^{2}$,
Eq.~\ref{eq:sig_dynamic}. This, however, has the disadvantage, that
correlations and fluctuations due to the the detector do not cancel
out. In order to remove the multiplicity dependence in the same fashion,
one can simply divide the dynamical fluctuations, $\sigma_{{\rm dynamical}}^{2}$,
by that of uncorrelated particles, $\sigma_{{\rm uncorrelated}}^{2}$,
Eq.~\ref{eq:variance_uncorr}, evaluated for the same number of particles.
\begin{equation}
f_{{\rm Poisson}}\equiv\frac{\sigma_{{\rm dynamical}}^{2}}{\sigma_{{\rm
uncorrelated}}^{2}}.\label{eq:possion}
\end{equation}
 with \[
\sigma_{{\rm uncorrelated}}^{2}=\frac{1}{\ave K}+\frac{1}{\ave{\pi}}\]
 for the case at hand. This scaling we will subsequently denote as
``Poisson'' scaling, as we scale with the scaled variance of a Poisson
distribution based on the observed multiplicities. The advantage of
this scaling is that it is unbiased in the sense that one does not
need to make any assumptions about the relative magnitude of the scaled
correlations. Another unbiased scaling would be to scale with the
number of particles involved, i.e. $N_{p}=\ave K+\ave{\pi}$. This
is similar in spirit of~\cite{Abelev:2009if} where a scaling with
the number of charged particles was studied. We shall henceforth refer
to this as particle number scaling,
\begin{equation}
f_{{\rm Particle\, Number}}=\left(\ave K+\ave{\pi}\right)\sigma_{{\rm
dynamical}}^{2}\label{eq:part_num_scaling}
\end{equation}
 Finally the expression for $\sigma_{{\rm dynamical}}^{2}$ in terms
of the scaled correlations, Eq.~\ref{eq:sig_dyn_scaled_corr}, suggests
the scaling with the kaon or pion number or with the geometric mean
of both, depending on which of the scaled correlations dominates.
\begin{eqnarray*}
f_{{\rm Kaon\, Number}} & = & \ave K\sigma_{{\rm dynamical}}^{2}\\
f{}_{\rm Pion\, Number} & = & \ave{\pi}\sigma_{{\rm dynamical}}^{2}\\
f_{{\rm geometric}} & = & \sqrt{\ave K\ave{\pi}}\sigma_{{\rm dynamical}}^{2}
\end{eqnarray*}
 Since the number of kaons is much smaller than the number of pions,
at least for the lower energies the kaon-number scaling is similar
to the Poisson scaling,
\[
\sigma_{{\rm uncorrelated}}^{2}=\frac{1}{\ave K}+\frac{1}{\ave{\pi}}=\frac{\ave
K+\ave{\pi}}{\ave K\ave{\pi}}\simeq\frac{1}{\ave K},
\]
 for $\ave K\ll\ave{\pi}$. Alternatively the above scaling relations
allow to relate the dynamical fluctuations at a given center of mass
energy with those at another energy. Specifically, given the dynamical
fluctuations at top RHIC energies, $\sqrt{s}=200\,{\rm GeV}$, the
dynamical fluctuations at any other center of mass energy is given
by 
\begin{itemize}
\item \emph{Poisson scaling:}\begin{equation}
\sigma_{{\rm dynamical}}\left(\sqrt{s}\right)=\sigma_{{\rm dynamical}}\left(200\,{\rm GeV}\right)\,\,\frac{\sqrt{\frac{1}{\ave K}+\frac{1}{\ave{\pi}}}|_{\sqrt{s}}}{\sqrt{\frac{1}{\ave K}+\frac{1}{\ave{\pi}}}|_{200\,{\rm GeV}}}\label{eq:poisson_scaling}\end{equation}

\item Particle Number scaling\begin{equation}
\sigma_{{\rm dynamical}}\left(\sqrt{s}\right)=\sigma_{{\rm dynamical}}\left(200\,{\rm GeV}\right)\,\,\frac{\sqrt{\ave K+\ave{\pi}}|_{200\,{\rm GeV}}}{\sqrt{\ave K+\ave{\pi}}|_{\sqrt{s}}}\label{eq:part_number_scaling_2}\end{equation}

\item $N_{K}$\emph{-scaling:} \begin{equation}
\sigma_{{\rm dynamical}}\left(\sqrt{s}\right)=\sigma_{{\rm dynamical}}\left(200\,{\rm GeV}\right)\,\,\frac{\sqrt{\ave K}|_{200\,{\rm GeV}}}{\sqrt{\ave K}|_{\sqrt{s}}}\label{eq:NK-scaling}\end{equation}

\item $N_{\pi}$\emph{-scaling:}\begin{equation}
\sigma_{{\rm dynamical}}\left(\sqrt{s}\right)=\sigma_{{\rm dynamical}}\left(200\,{\rm GeV}\right)\,\,\frac{\sqrt{\ave{\pi}}|_{200\,{\rm GeV}}}{\sqrt{\ave{\pi}}|_{\sqrt{s}}}\label{eq:npi_scaling}\end{equation}

\item \emph{Geometric Scaling:}\begin{equation}
\sigma_{{\rm dynamical}}\left(\sqrt{s}\right)=\sigma_{{\rm dynamical}}\left(200\,{\rm GeV}\right)\,\,\frac{\left(\ave K\ave{\pi}\right)^{1/4}|_{200\,{\rm GeV}}}{\left(\ave K\ave{\pi}\right)^{1/4}|_{\sqrt{s}}}\label{eq:geom_scaling}\end{equation}

\end{itemize}
The resulting \emph{scaled} dynamical fluctuations are shown in Fig.~\ref{fig:star_scale}
where we find that the energy dependence of $\sigma_{{\rm dynamical}}^{K/\pi}$
is reasonably reproduced by any of the scaling rules discussed above. The
values for the multiplicities of identified particles
taken from~\cite{Alt:2008ca,Abelev:2009if} as well as the result for the
rescaled fluctuations according to
Eqs.~(\ref{eq:poisson_scaling}-\ref{eq:geom_scaling}).
The essential point for the success of these scaling rules is that
we have used the number of \emph{identified} kaons and pions for the
mean values, $\ave K$ and $\ave{\pi}$, entering the scaling formulae,
Eqs.~\ref{eq:poisson_scaling}-\ref{eq:geom_scaling}. This leads to
an additional energy dependence especially for the NA49 data. Since
NA49 is a fixed target experiment, the actual acceptance and thus
the fraction of identified particles of the total number of particles
may vary considerably with the beam energy. This is illustrated in
Fig.~\ref{fig:star_scale_dndy}, where we applied the same scaling
formulae, Eqs.~\ref{eq:poisson_scaling}-\ref{eq:geom_scaling}, but
used the mid-rapidity yields, $\frac{{\rm d}N}{{\rm d}y}(y=0)$ for
the respective mean particle numbers, $\ave K$ and
$\ave{\pi}$~\cite{Abelev:2008ez,Alt:2007fe,Afanasiev:2002mx}. 
Obviously the energy dependence is not reproduced, especially for
the highest SPS energies, $\sqrt{s}=12.3\,{\rm and\,17.3\,{\rm GeV}}$.

Looking more closely at Fig.~\ref{fig:star_scale}, we see that the
STAR data at $\sqrt{s}=62.4\,{\rm {\rm GeV}}$, which have a rather
small error-bar are not well reproduced. Instead of arguing for new
physics in this energy regime, we note that in Ref.~\cite{Abelev:2009if}
the two most central values for $\sigma_{{\rm dynamical}}$ at this
energy do not agree very well with the systematics developed by the
STAR collaboration either. As already discussed in section~\ref{sec:Fluctuations-of-Particle},
since $\ave K/\ave{\pi}\ll1$ the dynamical fluctuations are dominated
by the kaons. Therefore, the kaon-number or Poisson scaling should
work better than pion-number or particle number scaling. This is indeed
the case. 

One may be tempted to use the quality of agreement of the various
scaling prescriptions to draw conclusions about the importance of
the various contributions to the dynamical fluctuations. Given the
experimental error-bars and the quality of agreement of the various
scaling rules the value of such an exercise is not obvious.
At least,
a global fit based on the various scaling prescriptions needs to be
carried out before any more detailed conclusions about the strength
of the various contributions can be drawn. 

Finally let us comment on the fact that calculations based on the
UrQMD model do not reproduce the rise of the dynamical fluctuations
as observed by NA49, although the NA49 acceptance has been applied
\cite{Alt:2008ca}. First of all, while UrQMD seems to do reasonably
well for mid-rapidity abundances, it is not clear if UrQMD does reproduce
the correct multiplicity of identified particles \emph{within} the
acceptance,
especially at small center of mass energies. Second, it is not at
all obvious, if UrQMD contains indeed all the relevant sources for
fluctuations and correlations. For instance quantum statistics is
not taken into account. The fact that the above scaling relations
are able to connect the dynamical fluctuations over a wide range of
beam energies does not say anything about the nature and origin of
these fluctuations except that they seem to be the same at all energies.
While this considerably weakens the argument for the $K/\pi$-fluctuations
being a signature for the QCD critical point, the disagreement with
UrQMD may very well point to a yet to be discovered new source for
fluctuations/correlations. This however, requires that all {}``trivial''
sources, such as e.g.\ quantum statistics, are systematically taken
into account.

\begin{figure}[ht]
\includegraphics[width=0.7\textwidth]{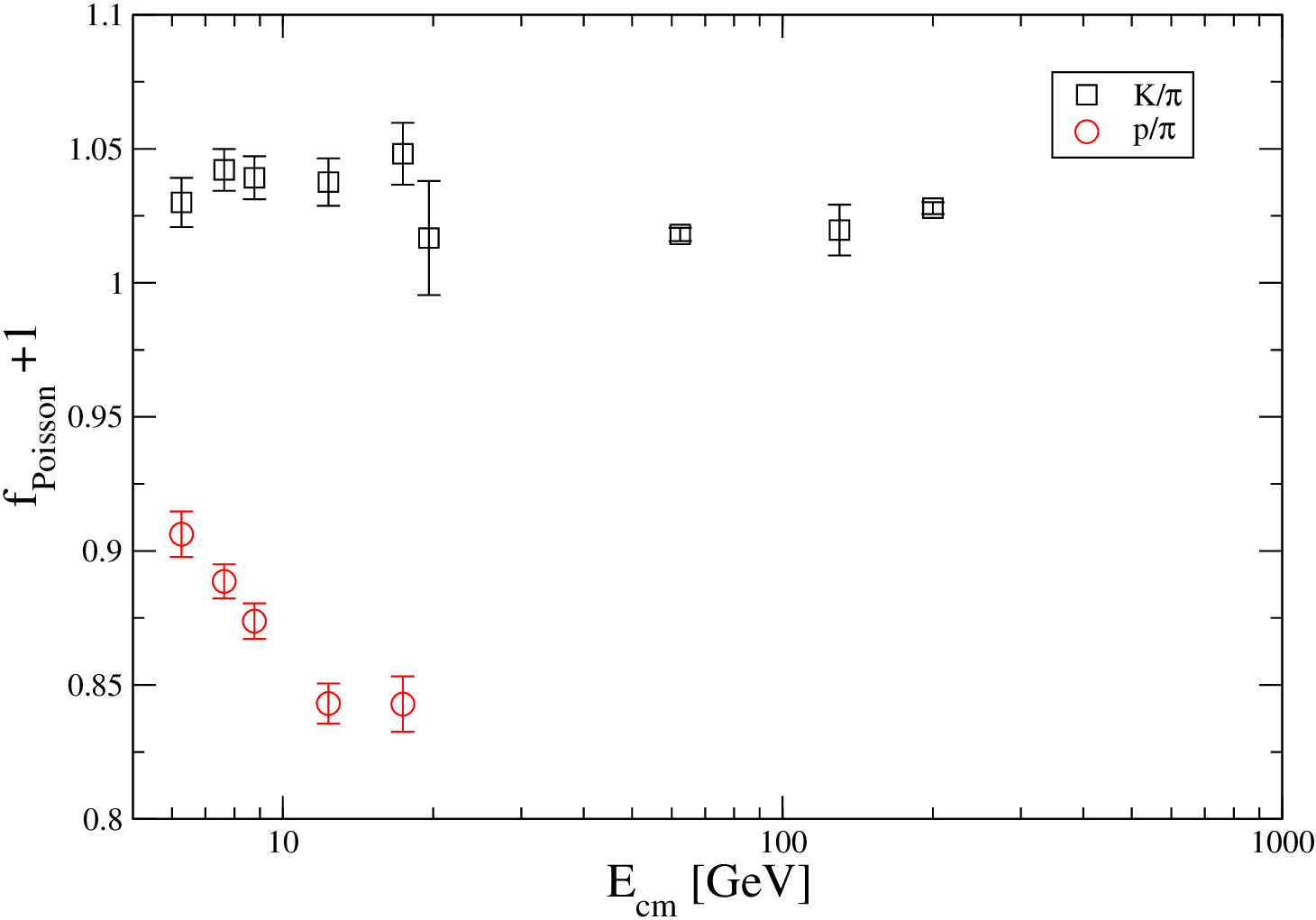}

\caption{The rescaled fluctuations $f_{{\rm Poisson}}+1$ for $K/\pi$ (black)
and $p/\pi$-fluctuations (red).}

\label{fig:f_poisson+1} 
\end{figure}

Finally, let us conclude this section by proposing an observable,
which should, to leading order, be independent of the multiplicity.
As already discussed at the beginning of this section, the ratio \[
f\equiv\frac{\sigma^{2}}{\sigma_{{\rm mixed}}^{2}}\]
 proposed in~\cite{Jeon:1999gr} has the disadvantage that correlations
and fluctuations due to the detector do not cancel out. We, therefore,
propose to study instead the ratio, Eq.~\ref{eq:possion} 
\[
f_{{\rm Poisson}}\equiv\frac{\sigma_{{\rm dynamical}}^{2}}{\sigma_{{\rm
uncorrelated}}^{2}}
\]
or equivalently $f_{{\rm Poisson}}+1$, which has the same limit
as $f$ in the absence of correlations. In Fig.~\ref{fig:f_poisson+1}
we show $f_{{\rm Poisson}}+1$ for both the $K/\pi$ fluctuations
as well as for the $p/\pi$-fluctuations, which have been measured
by NA49 as well. We see a rather weak energy dependence of the
$K/\pi$-fluctuations, whereas the $p/\pi$-fluctuations exhibit a
variation with energy. We note, however, that the energy dependence
of the $p/\pi$-fluctuations in the scaled variable is opposite to
that in $\sigma_{{\rm dynamical}}^{2}.$ While $\sigma_{{\rm dynamical}}^{2}$
decreases with energy, the magnitude of $f_{\rm Poisson}$ increases,
suggesting in increase of the the strength of the correlations with
energy, which seems to level off at top SPS energies. This suggests
that the correlations, mostly due to baryon resonances, increase with
energy up to $\sqrt{s}\simeq15\,{\rm GeV}$. Using the chemical freeze-out
parameters of Ref.~\cite{Braun-Munzinger:2003zd} and assuming that
only the delta resonance contributes to the scaled correlation coefficient
$C_{p\pi}$ we find indeed an increase of the correlations from AGS
up to SPS energies. 

\begin{table}[ht]
\begin{tabular}{|c|c|c|c|c|c|c|c|c|c|}
\hline 
$\sqrt{s}${[}GeV{]}  & $\ave{\pi}_{ident.}$  & $\ave{K}_{ident.}$  &
$\sigma_{{\rm dyn.}}[\%]$ &
Geom. scaling  & $N_{K}$-scaling  & $N_{\pi}$-scaling  & Poisson-scaling & Part.
Num. scaling & $f_{\rm Poisson}$(\%)\tabularnewline
\hline
\hline 
6.27  & 30.9  & 5.7  & 7.89 & 10.05  & 7.15 & 14.11 & 7.61 & 13.27 &
3.0\tabularnewline
\hline 
7.63  & 66.6  & 10.2  & 6.90 & 7.17  & 5.35 & 9.60 & 5.61 & 9.15 &
4.2\tabularnewline
\hline 
8.77  & 103.3  & 15.0  & 5.47  & 5.83 & 4.41 & 7.71 & 4.61 & 7.37 &
3.9\tabularnewline
\hline 
12.3  & 227.6 & 31.2  & 3.70  & 3.99 & 3.05 & 5.19 & 3.18 & 4.98 &
3.8\tabularnewline
\hline 
17.3  & 416.6 & 54.2  & 3.17  & 2.99 & 2.32 & 3.84 & 2.41 & 3.69 &
4.8\tabularnewline
\hline 
19.6 & 227.6 & 10.7 & 4.04 & 5.21 & 5.21 & 5.19 & 5.21 & 5.19 &
1.7\tabularnewline
\hline 
62.4  & 319.2 & 14.5  & 3.61  & 4.44 & 4.48 & 4.38 & 4.48  & 4.39 &
1.8\tabularnewline
\hline 
130 & 351.9 & 14.4 & 3.78 & 4.34 & 4.51 & 4.18 & 4.49 & 4.19 &
2.0\tabularnewline
\hline 
200  & 432.6  & 20.6  & 3.77  & 3.77  & 3.77 & 3.77 & 3.77 & 3.77 &
2.8\tabularnewline
\hline
\end{tabular}

\caption{Table of the results for the various scaling scenarios
Eqs.~(\ref{eq:poisson_scaling}-\ref{eq:geom_scaling})
using the STAR data at $\sqrt{s}=200\,{\rm GeV}$ as reference.
The data for identified multiplicities, $\ave{\pi}_{ident.}$ and
$\ave{K}_{ident.}$, are from
\cite{Alt:2008ca,Abelev:2009if}.}

\label{tab:star_fit} 
\end{table}

\section{Conclusions}

In this article we have reviewed the multiplicity dependence of particle
ratio fluctuations. We have provided several scaling prescriptions
which correct for the inherent dependence of the dynamical fluctuations
$\sigma_{{\rm dynamical}}^{2}$ on the number of \emph{identified}
particles. We have demonstrated that these scaling rules naturally
reproduce the trend seen in the energy dependence of the kaon-to-pion
fluctuations. Consequently any interpretation of the rise of $\sigma_{{\rm dynamical}}^{K/\pi}$
towards lower energies in the context of a possible QCD critical point,
needs to account for the {}``trivial'' effect due to the multiplicity
dependence. We propose that future studies of energy and / or multiplicity
dependencies should correct for the {}``trivial'' multiplicity dependencies
inherent in $\sigma_{\rm dynamical}$. In our view, the simplest and least
biased scaling is the Poisson scaling,
$f_{\rm Poisson}$, which could serve as a benchmark.
Of course a more complete analysis of the energy dependence would
include a thorough global fit of the available data, which we have
not carried out in this paper.

We have further applied the scaling to the $p/\pi$-fluctuations measured
by the NA49 collaboration. We find that the properly scaled observable
still exhibits a strong energy dependence, which, however, is opposite
to the unscaled one. In this context it would be interesting to include
the STAR data, which unfortunately are not available in a final version
yet.

Finally we point out that any multiplicity scaling needs to be based
on the mean multiplicity of the actual identified particles used for
the fluctuation measurement instead of an extrapolated multiplicity
such as ${\rm d}N/{\rm d}y$.


\begin{acknowledgments}
The authors would like to thank G. Westfall and R. Stock for useful
discussions. 
This work was initiated during the Program {}``The QCD Critical Point''
at the Institute for Nuclear Theory (INT) in Seattle, WA. The authors
would like to thank the INT for the hospitality during this program.
This work is supported by the Director, Office of Energy Research,
Office of High Energy and Nuclear Physics, Divisions of Nuclear Physics,
of the U.S. Department of Energy under Contract No. DE-AC02-05CH11231.
The authors also acknowledge support by the Deutsche Forschungsgemeinschaft
(DFG) and the Helmholtz Research School on Quark Matter Studies.


\end{acknowledgments}

\end{document}